\documentclass[useAMS,usenatbib]{mn2e}
\usepackage{graphicx} 
\usepackage{times}
\usepackage{epsfig}
\usepackage{rotating}
\usepackage{sidecap}
\usepackage{longtable}
\usepackage{lscape}
\usepackage{amssymb}

\title[The Relativistic precession model]{Solutions to the relativistic
  precession model}
\author[A. Ingram \& S.E. Motta]{Adam
  Ingram$^{1}\thanks{E-mail:a.r.ingram@uva.nl}$ \& Sara Motta$^{2}$\\
$^1$Astronomical Institute, Anton , University of Amsterdam, Science Park 904, 1098 XH Amsterdam, the Netherlands.\\
$^2$ESAC, European Space Astronomy Centre, Villanueva de la Ca\~nada, E-28692 Madrid, Spain\\}

\begin{document}
\date{Accepted 2014 August 4. Received 2014 July 4; in original form 2014 July 4}

\pagerange{\pageref{firstpage}--\pageref{lastpage}} \pubyear{2014}

\maketitle
\begin{abstract}

The relativistic precession model (RPM) can be used to obtain a precise
measurement of the mass and spin of a black hole when the appropriate set
of quasi periodic oscillations is detected in the power-density
spectrum of an accreting black hole. However, in previous studies the solution of the
RPM equations could be obtained only through numerical methods at a price
of an intensive computational effort. Here we demonstrate that the
RPM system of equations can be solved analytically, drastically
reducing the  computational load, now limited to the Monte-Carlo
simulation necessary to estimate the uncertainties. The analytical
method not only provides an easy solution to the RPM system when three
oscillations are detected, but in all the cases where the detection of
two simultaneous oscillations is coupled with an independent mass
measurement. We also present a computationally inexpensive method to
place limits on the black hole mass and spin when only two oscillations are
observed.
\end{abstract}


%===============================================
\section{Introduction}
\label{sec:intro}
%===============================================

Quasi periodic oscillations (QPOs) were discovered several decades ago
in the X-ray flux of accreting stellar mass black holes (BHs) and
neutron stars (NSs). It is now clear that QPOs are a common
characteristic of accreting systems, having also been observed from
\textit{Ultra Luminous X-ray sources} (\citealt{Strohmayer2003}) and,
for a few cases, Active Galactic Nuclei (AGN,
\citealt{Gierlinski2008}). QPOs take the form of narrow peaks in the
Fourier power spectrum of the X-ray light curve, and thus their
centroid frequencies can be measured with high accuracy, offering the
opportunity to accurately probe the distorted spacetime in the
vicinity of a compact object. From their short timescales and high
coherence, simple light crossing arguments indicate that these
phenomena must originate from the innermost regions of the accretion
flow.

In spite of being studied extensively since their discovery, the
physical origin of QPOs remains ambiguous. However, years of
comprehensive monitoring by the Rossi X-ray Timing Explorer
(\textit{RXTE}) has yielded a detailed phenomenological knowledge of
QPO observational properties. In BH X-ray binaries, low frequency QPOs
(LF QPOs) are very strong and commonly observed features and  have been
split up into three subclasses: Type-A, B and C (see
e.g. \citealt{Wijnands1999}, \citealt{Casella2005},
\citealt{Motta2012}). Type-C QPOs are by far the most commonly
observed. Their centroid frequency usually varies in the  $\sim
0.1$--30 Hz range and tightly correlates with the spectral 
evolution of the host source (see e.g. \citealt{Belloni2011}). 
Pairs of high frequency QPOs (HF QPOs), with centroid frequencies
$\gtrsim 100$ Hz, have also been observed, even though they are much
harder to detect above the Poisson noise level.  Nonetheless, they
have sparked much theoretical interest because their frequencies are
commensurate with the orbital frequency close to the BH (see
e.g. \citealt{Abramowicz2001}, \citealt{Kluzniak2001},
\citealt{Lamb2001}). LF QPOs are also observed in NS X-ray binaries
with higher centroid frequencies, consistently with linear mass
scaling. The NS analogy to HF QPOs are kHz QPOs which, in contrast to
their BH counterparts, are regularly observed, often with very high
amplitude (\citealt{vanderKlis1996a}).

There are many suggested QPO mechanisms in the literature that can be divided into two main groups: those associated to wave modes of the accretion flow (\citealt{Tagger1999}, \citealt{Titarchuk1999}, \citealt{Wagoner2001}, \citealt{Cabanac2010}), and those associated with relativistic effects that involve the misalignment  of the accretion flow and the black hole spin (\citealt{Stella1998}, \citealt{Lamb2001}, \citealt{Abramowicz2001}, \citealt{Fragile2005}, \citealt{Schnittman2006}, \citealt{Homan2006}, \citealt{Ingram2011}). This second group of models are based on the idea that, whereas in Newtonian gravity bound elliptical orbits around a point-like gravitating mass always remain in the same plane with a stationary semi-major axis, in the theory of General Relativity (GR) this is not the case. Mathematically, this means that the three coordinate frequencies: orbital, vertical and radial epicyclic are not equal, $\nu_{\rm \phi} \neq \nu_{\rm \theta} \neq \nu_{\rm r}$. Periastron precession, with frequency $\nu_{\rm per} = \nu_{\rm \phi} - \nu_{\rm r}$ is a precession of an elliptical orbit's semi-major axis. Nodal (Lense-Thirring) precession, which occurs only for orbits out of the equatorial plane of a spinning gravitating mass, is a precession of the orbit's spin axis around the spin axis of the gravitating mass. This has a frequency $\nu_{\rm nod} = \nu_{\rm \phi} - \nu_{\rm \theta}$. All of these frequencies depend only on the radius of the orbit, $r$, and the mass, $M$, and dimensionless spin parameter, $-1<a<1$, of the gravitating mass. 

In the relativistic precession model (RPM) proposed by \citealt{Stella1998}, the Type-C QPO originates from nodal precession, the lower HF QPO from periastron precession and the upper HF QPO from orbital motion, with all three signals originating from one characteristic radius, $r$. The inward movement of this radius can then explain the observed co-evolution of the three QPOs to higher frequencies (e.g. \citealt{Stella1999}). 
This model has been applied with mixed success to NSs
(\citealt{Stella1999a}; \citealt{Ingram2010};
\citealt{Altamirano2012a}), but appears to work very well for BHs. In
particular, Motta et al. (2014; hereafter M14) considered an
observation of GRO 1655-40 in which the presence of three
simultaneously observed QPOs leaves three equations and three
unknowns. They were thus able to solve the equations of the RPM to
obtain values of $r$, $a$ and $M$. Encouragingly, they found that the
mass measured in this way agrees very well with the dynamical mass
measurement for this source (\citealt{Beer2002}). Unfortunately, the
simultaneous occurrence of the three QPOs relevant for the RPM is
extremely rare and so it is not possible to simply apply this
technique to every BH. However, a spin measurement can be achieved
also in the case where the mass of the BH is known and two
simultaneous QPOs are detected, as suggested by Motta et al (2014a;
hereafter M14a).

In previous studies, the equations of the RPM have been solved
numerically in a very computationally intensive manner (M14; M14a;
\citealt{Bambi2014}; \citealt{2014arXiv1403.4136S}), resulting from a
belief that the equations cannot be solved analytically. Here, we
present an analytic solution for the case where three QPOs are
detected. We also present simple methods to solve for all occurrences
where two QPO detections are combined with a mass
measurement. Although we could not find an entirely analytic solution
for the latter case, our method is far quicker than all previously
used methods. In addition we consider a method to place tight limits on
system parameters with only two QPOs and no mass measurement.

%===============================================
\section{Solving the system with three QPOs}
\label{sec:solution}
%===============================================

In the case of a test mass orbiting a spinning BH in a plane slightly
perturbed from equatorial, it can be shown that, in Kerr metric
(\citealt{Bardeen1972}; \citealt{Merloni1999}), the
orbital, periastron precession and nodal precession frequencies are
given by:

\begin{eqnarray}
\nu_{\rm \phi} & = & \pm \frac{\beta}{M} \frac{1}{r^{3/2} \pm a} \label{eqn:nuphi} \\
\nu_{\rm per} & = & \nu_{\rm \phi} \left[  1 - \sqrt{ 1 - \frac{6}{r}
    \pm \frac{8a}{r^{3/2}} - \frac{3a^2}{r^2} }
\right] \label{eqn:nuper} \\
\nu_{\rm nod} & = & \nu_{\rm \phi} \left[  1 - \sqrt{ 1 \mp
    \frac{4a}{r^{3/2}} + \frac{3a^2}{r^2} } \right], \label{eqn:nunod} 
\end{eqnarray}
where $M$ is BH mass in units of solar masses, $r$ is radius in units
of $R_g=GM M_\odot/c^2$, $a$ is the dimensionless spin parameter and
$\beta=c^3/( 2 \pi G M_\odot )=3.237 \times 10^4$ Hz. In all equations, the top sign refers to prograde spin (i.e. orbital
motion is in the same direction as BH spin) and the bottom sign refers to
retrograde spin. 
Since no stable orbits exist inside of the innermost
stable circular orbit (ISCO), we can set the extra condition $r > r_{\rm
  ISCO}$. The radius $r_{\rm ISCO}$ depends monotonically on the spin (see \citealt{Bardeen1972}; M14), ranging from $9>r_{\rm ISCO}>1$ for $-1<a<1$ and taking the
value $r_{\rm ISCO}=6$ for $a=0$.

In the RPM, $\nu_{\rm LF}=|\nu_{\rm nod}|$, $\nu_{l} = |\nu_{\rm per}|$
and $\nu_{\rm u} = |\nu_{\rm \phi}|$, where $\nu_{\rm LF}$, $\nu_{\rm
  l}$ and $\nu_{\rm u}$ are respectively the measured Type-C, lower HF
and upper HF QPO frequencies (M14). The equations of the RPM depend on
the mass and the spin of the compact object, on the radius at which
the frequencies are produced and on the frequencies themselves. If we
have measurements of all three QPO frequencies simultaneously, we can
solve for the three remaining unknowns (mass, spin and emission
radius), assuming that all the frequencies are associated with the
same radius.

We see that the mass is explicitly contained only in equation
\ref{eqn:nuphi} and so the equations for $\nu_{\rm per}$ and $\nu_{\rm
  nod}$ form a system of two simultaneous equations which we can solve
to get $r$ and $a$ before calculating $M$ from equation
\ref{eqn:nuphi}. 
Re-arranging equations \ref{eqn:nuper} and
\ref{eqn:nunod} gives:
\begin{eqnarray}
\Gamma & \equiv & \left( 1  - \frac{\nu_{\rm per}}{\nu_{\rm \phi}} \right)^2
= 1 - \frac{6}{r} \pm \frac{8a}{r^{3/2}} - \frac{3a^2}{r^2}
\label{eqn:Gamma} \\
\Delta & \equiv &  \left( 1  - \frac{\nu_{\rm nod}}{\nu_{\rm \phi}} \right)^2
= 1 \mp \frac{4a}{r^{3/2}} + \frac{3a^2}{r^2}.
\label{eqn:Delta}
\end{eqnarray}
For $r > r_{\rm ISCO}$, these constants obey $0<\Delta <1$, $0<\Gamma <1$,
and $\Delta > \Gamma$. Adding together equations \ref{eqn:Gamma} and
\ref{eqn:Delta} gives $a$ in terms of $r$:
\begin{equation}
a = \pm \frac{r^{3/2}}{4} \left[ \Delta + \Gamma - 2 + \frac{6}{r} \right].
\label{eqn:a}
\end{equation}
Taking twice equation \ref{eqn:Delta} and adding it to
\ref{eqn:Gamma} gives:
\begin{equation}
3 - 2\Delta - \Gamma - \frac{6}{r} + \frac{3a^2}{r^2} = 0.
\label{eqn:a2}
\end{equation}
Substituting equation \ref{eqn:a} into \ref{eqn:a2} and re-arranging
(including multiplying by $r$) gives a quadratic in $r$:
\begin{equation}
\frac{3}{4}(\Delta+\Gamma-2)^2r^2 + (\Delta+5\Gamma - 6)r + 3 = 0,
\label{eqn:quad}
\end{equation}
which can be solved using the quadratic formula. After re-arranging,
this gives the solution for $r$,
\begin{equation}
r = \frac{2}{3} \frac{6-\Delta-5\Gamma + 2
  \sqrt{2(\Delta-\Gamma)(3-\Delta-2\Gamma)} }
{(\Delta+\Gamma-2)^2}.
\label{eqn:r}
\end{equation}
From this, the spin can be determined from equation \ref{eqn:a} and
the mass by re-arranging equation \ref{eqn:nuphi}.

Note that there is a degeneracy between pro-grade and retro-grade
spin: although there are two solutions to the quadratic in $r$, the
alternate solution with a minus sign before the determinant is a
solution to the set of equations \ref{eqn:Gamma} and \ref{eqn:Delta},
but \textit{not} to the set of equations \ref{eqn:nuper} and
\ref{eqn:nunod}; which are the equations we actually want to solve
for. The solutions for mass and radius are identical with the spin
being $\pm$ the value derived assuming prograde motion. This means we
can derive $r$, $M$ and $|a|$ assuming prograde spin, but we do not
know if the spin is prograde or retrograde. This degeneracy can be
broken by measuring the highest frequency reached by the Type-C QPO:
if it extends to within the ISCO for $a=-|a|$, we can assume prograde
spin.

Since these solutions for $r$, $a$ and $M$ are all differentiable, it
would in principle possible to apply the standard error propagation to
determine uncertainties. However, these solution are non linear
functions of $r$, $a$ and $M$, therefore the standard error
propagation formula is not appropriate. Nevertheless, error estimates
can be obtained through a Monte-Carlo simulation following the method
outlined in M14. For each step, values for $\nu_{\rm nod}$, $\nu_{\rm per}$
and $\nu_\phi$ are chosen from Gaussian distributions with mean
$\nu_{\rm LF}$, $\nu_{l}$ and $\nu_{\rm u}$, respectively, and standard
deviation $d\nu_{\rm LF}$, $d\nu_{l}$ and $d\nu_{\rm u}$,
respectively. Solutions for $r$, $a$ and $M$ can then be obtained
analytically for each step. The mean and standard deviation for each
of these three quantities then give the measurement and
error. This process was very time consuming for the cases of M14 since
at each step the RPM equations were solved numerically, but is very
fast using the analytic formulae presented here. Of course, applying
this method to the three QPOs found in GRO J1655-40, we obtained the
same solution of the RPM presented in M14 (see
Tab. \ref{tab:solutions}).

%===============================================
\section{Solving the system with two QPOs and a mass measurement}
\label{sec:solution2}
%===============================================

Detection of three simultaneous QPOs is very rare for a BH. In fact,
the case of GRO J1655-40 considered by M14 is so far the only reported
occurrence. There are, however, detections of two simultaneous
QPOs in objects which have a reliable dynamical mass measurement. This
is the case for XTE J1550-564, which displays simultaneously a Type-C
LF QPO and a lower HF QPO (M14a). Here, we present computationally
inexpensive solutions for detections of all three possible
combinations of two simultaneous QPOs coupled with a mass
measurement.

\subsection{$\nu_{\rm per}$ is the unknown}\label{sec:nuper}

In the case where we have no measurement of $\nu_{\rm per}$, our three
unknowns are $a$, $r$ and $\nu_{\rm per}$, while  $M$, $\nu_{\rm nod}$
and $\nu_{\rm \phi}$ are known. We can express the spin as a function
of only one unknown ($r$) by re-arranging equation \ref{eqn:nuphi} to
get:
\begin{equation}
a = \Theta \mp r^{3/2},
\label{eqn:Sigma}
\end{equation}
where $\Theta \equiv \beta / (\nu_{\rm \phi} M)$. 
Combining this with equation \ref{eqn:Delta} gives:
\begin{equation}
3r^3 + (5-\Delta)r^2 \mp 6\Theta r^{3/2} \mp 4\Theta r^{1/2} + 3\Theta^2 =0.
\end{equation}
Using the substitution $r=x^2$ leaves us with a 6$^{\rm th}$
order polynomial:
\begin{equation}
3x^6 + (5-\Delta)x^4 \mp 6\Theta x^3 \mp 4\Theta x + 3\Theta^2 =0.
\label{eqn:x}
\end{equation}
Unfortunately, we were unable to find an analytic solution to this
equation, but since it is a polynomial, all the roots can be found
using Laguerre's method. 
We find all six complex roots using the code
\textsc{zroots} (\citealt{Press1992}) and find that, for all parameter
combinations trialled, there is only one real root for $x$ (and
therefore for $r$; i.e. the other five roots are always complex). 
The spin can then be calculated from equation \ref{eqn:Sigma}. 
We stress that the process of solving equation \ref{eqn:x} using Laguerre's
method is far quicker than solving the entire set of three
simultaneous equations using Newton's method as in previous works.

Curiously, we find that the one real root of equation \ref{eqn:x} is
\textit{independent} of whether we assume prograde or retrograde
spin. The equation expressed using the top signs shares a common root
with the equation using the bottom signs, and this happens to be the
only real root. We can thus always assume prograde spin for the
purposes of finding a solution for $r$ and $|a|$ (and $\nu_{\rm
  per}$). The spin could then be $\pm |a|$ and we can again attempt to
break the degeneracy by assessing whether orbits pass inside the ISCO
for the retrograde solution.

We applied this procedure to GRO J1655-40, in the two cases where a
type-C QPO was detected simultaneously to a upper HF QPO (M14). We used
the value of the mass obtained from spectro-photometric observations
and we obtained the spin measurements reported in
Table\ref{tab:solutions}, which are consistent with the ones reported
in M14.

\subsection{$\nu_{\rm nod}$ is the unknown}\label{sec:nunod}

By far the easiest QPO to detect is the Type-C LF QPO (\citealt{Motta2012}) which is associated
with $\nu_{\rm nod}$ in the RPM. However, there are some cases where
the two HF QPOs are detected, but not the type-C QPO
(e.g. \citealt{Homan2005}). Therefore, we also present the equation
for $r$ (analog to equation \ref{eqn:x}) in the case where we measure
the frequency of two HF QPOs, but no LF QPO. It is derived by
combining equation \ref{eqn:Sigma} with equation \ref{eqn:Gamma}
\begin{equation}
3x^6 + (7+\Gamma)x^4 \mp 6\Theta x^3 + 6x^2 \mp 8\Theta x + 3\Theta^2 =0.
\label{eqn:x2}
\end{equation}
Similarly to the case described in Sec. \ref{sec:nuper}, this equation
can be solved using Laguerre's method to get a solution for $r$ and
$|a|$ for prograde and retrograde spin.

\subsection{$\nu_{\rm \phi}$ is the unknown}\label{sec:nuphi}

If only a Type-C QPO and the lower HF QPO are detected (as in the case
of XTE J1550-564, M14a), we cannot find an analytic solution. However,
we can solve the system numerically with very little computational
expense if we attempt to solve for $\nu_{\rm \phi}$ rather than for
$M$. 

\begin{figure}
\begin{centering}
 \includegraphics[height=8cm,width=8cm,trim=0.0cm 0.0cm 0.0cm
 0.0cm,clip=true]{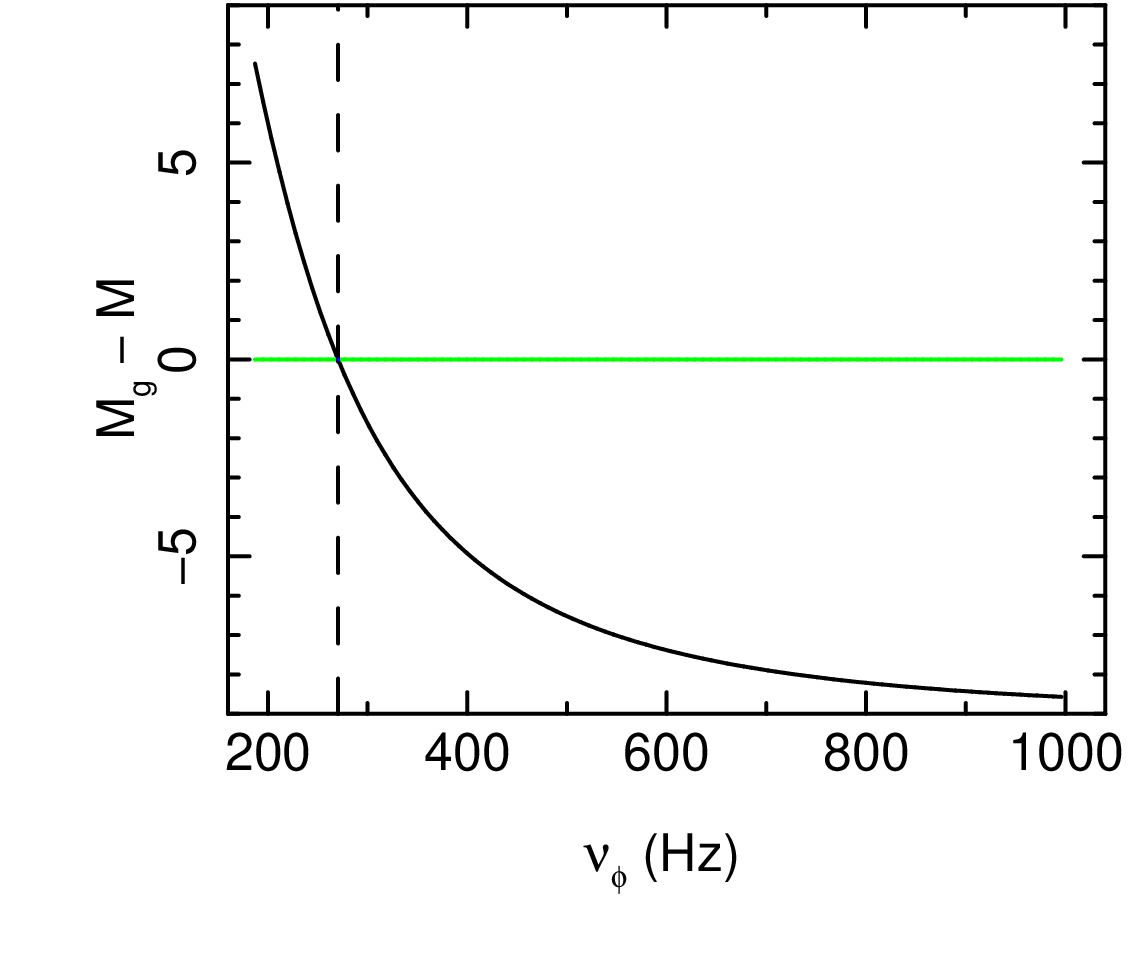}
\end{centering}
\vspace{-0.5cm}
 \caption{Function $f(\nu_{\rm \phi} , \nu_{\rm per} ,\nu_{\rm nod} ,
   M ) =  M_{\rm g} - M$ for the case of XTE 1550-564 (M14a), when
   $\nu_{\rm nod}$, $\nu_{\rm per}$ and $M$ are known. The solution
   for $\nu_{\rm \phi}$ corresponds to the point where this function
   crosses $M_{\rm g} - M = 0 $  (green line). This gives
   $\nu_\phi=270.5$ Hz}
 \label{fig:nuphi}
\end{figure}

We know $\nu_{\rm nod}$, $\nu_{\rm per}$ and $M$. If we make a
guess for $\nu_{\rm \phi}$, we can then
calculate a guess for $r$ from equation \ref{eqn:r}. From this we can
calculate a guess for $a$ using equation \ref{eqn:a} and finally
calculate a guess for the mass, $M_{\rm g}$, by re-arranging equation
\ref{eqn:nuphi}. The black line in Figure \ref{fig:nuphi} shows the function:

\begin{equation}
f(\nu_{\rm \phi} , \nu_{\rm per} ,\nu_{\rm nod} , M ) =  M_{\rm g} - M,
\end{equation}
plotted against $\nu_{\rm \phi}$, assuming $\nu_{\rm nod}=13.08$ Hz,
$\nu_{\rm per} = 183$ Hz and $M=9.1$, as is the case for XTE
J1550-564 (M14a). The solution for $\nu_{\rm \phi}$ occurs
when this function crosses zero (marked by the green line). Since this
is a well behaved function, it is very simple and robust to find a
solution using the bisection method (we use \textsc{rtbis} from \cite{Press1992} and assume $\nu_{\rm per} < \nu_{\rm \phi} < 1000$ Hz).

%%Left, Bottom, Right, Top

This procedure yields a solution $\nu_{\rm \phi} = 270.5$ Hz. The
radius and spin can then be determined respectively from equations
\ref{eqn:r} and \ref{eqn:a} to give $r=5.476$ and $a=0.339$,
consistent with the values reported by M14a (see Table
\ref{tab:solutions} for error estimates).

%===============================================
\section{Placing limits with two QPOs and no independent mass measurement}
\label{sec:solution3}
%===============================================

Even if we do not have an independent mass measurement, we can still
place limits on the system by assuming that we do not see orbits
inside of the ISCO, in a manner similar to
\cite{2014arXiv1403.4136S}. This means that 
the highest frequency Type-C QPO we observe must come from a radius
larger than or equal to $r_{\rm ISCO}$. For the case of XTE J1550-564 which has a dynamical
mass measurement of $M=9.1$, M14a showed that $r_{\rm ISCO}=4.83$ for
their spin measurement of $a=0.341$ and that the nodal precession
frequency at $r_{\rm ISCO}$ for this spin and mass is $\nu_{\rm nod}({\rm ISCO}) =
18.8$ Hz. This is encouraging since the highest frequency Type-C QPO
ever observed from this source has a frequency $\nu_{\rm LF}({\rm
  max})=18.04$ Hz. The data are thus consistent with the requirement
of the model that $\nu_{\rm nod}({\rm ISCO}) \geq \nu_{\rm LF}({\rm
  max})$.

\begin{figure}
\begin{centering}
 \includegraphics[height=8cm,width=8cm,trim=0.0cm 0.0cm 0.0cm
 0.0cm,clip=true]{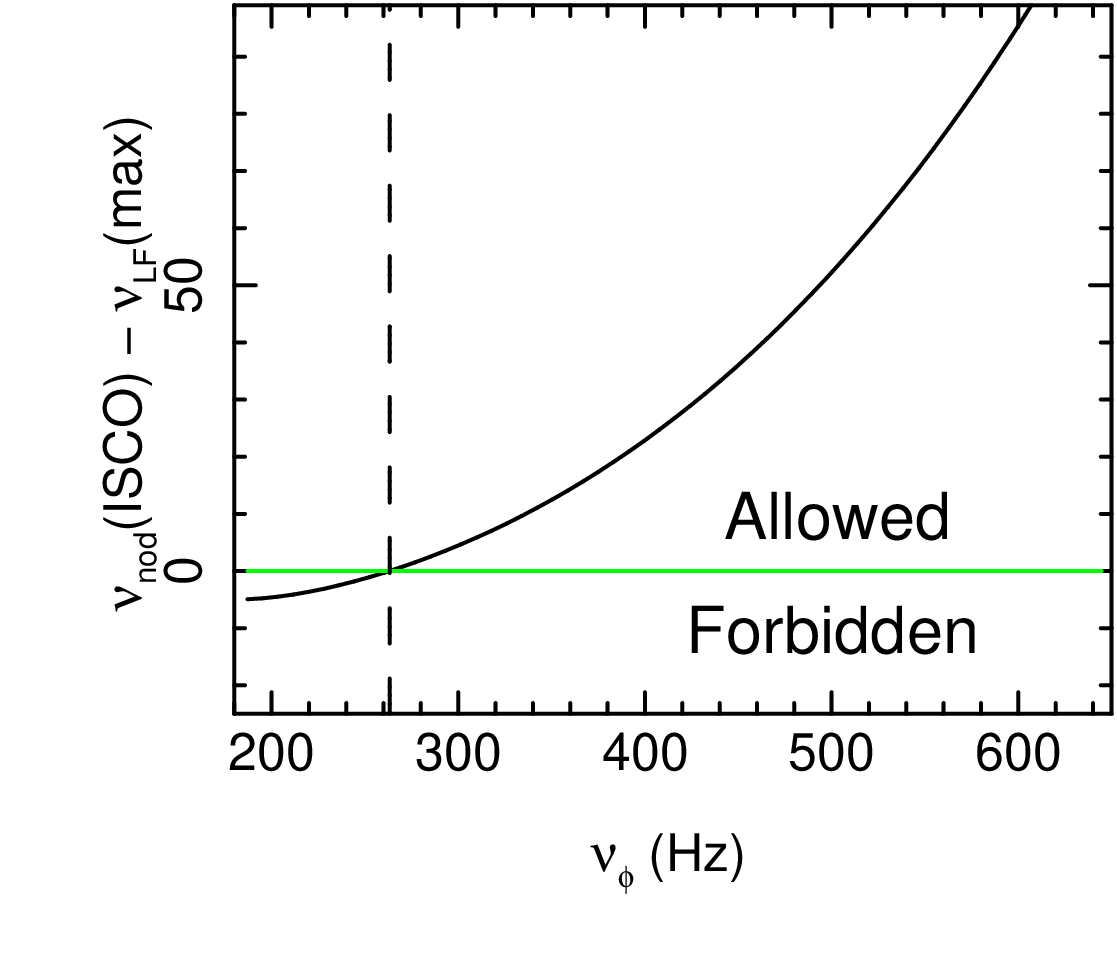}
\end{centering}
\vspace{-0.5cm}
 \caption{Function $f(\nu_{\rm \phi} , \nu_{\rm per} ,\nu_{\rm nod} , \nu_{\rm LF}({\rm max}) ) =
\nu_{\rm nod}({\rm ISCO}) - \nu_{\rm LF}({\rm max})$ for the case of
XTE J1550-564, when only $\nu_{\rm nod}$ and $\nu_{\rm per}$ are
known. The intersection between the black line and the green line
(that marks $\nu_{\rm nod}({\rm ISCO}) - \nu_{\rm LF}({\rm max}) = $
0) corresponds to the lower limit on $\nu_{\rm \phi}$ ($\nu_\phi \geq
263$ Hz.}
 \label{fig:ISCOtest}
\end{figure}

We can use this reasoning to place limits on the system
\textit{without} an independent mass measurement. 
Let us consider the case of XTE J1550-564, but say we do not have a mass measurement. We 
have a simultaneous measurement of $\nu_{\rm nod}$ and $\nu_{\rm per}$
and we also have a measurement of $\nu_{\rm LF}({\rm max})$. We can
apply the same trick as in section \ref{sec:nuphi}: we make a guess for
$\nu_{\rm \phi}$ and from that calculate guesses for $r$, $a$ and $M$
using equations \ref{eqn:r}, \ref{eqn:a} and \ref{eqn:nuphi}. From
this, we can calculate a guess for $r_{\rm ISCO}$ and finally a guess
for $\nu_{\rm nod}({\rm ISCO})$. In Figure \ref{fig:ISCOtest}, we plot
the function:
\begin{equation}
f(\nu_{\rm \phi} , \nu_{\rm per} ,\nu_{\rm nod} , \nu_{\rm LF}({\rm max}) ) =
\nu_{\rm nod}({\rm ISCO}) - \nu_{\rm LF}({\rm max}),
\label{eqn:ISCOtest}
\end{equation}
against $\nu_{\rm \phi}$. We can find a lower limit on $\nu_{\rm
  \phi}$ by determining where this function crosses zero (again, we
use the bisection method). 
For XTE J1550-564, we find $\nu_{\rm \phi}
\geq 263$ Hz. From this, we can use the equations in section
\ref{sec:solution} to find $r \geq 5.39$, $a \leq 0.341$ and $M \leq
9.56$ (see Tab. \ref{tab:solutions}). Since the \textit{RXTE} monitoring of these sources was very
comprehensive, it is likely that we should be able to find a Type-C
QPO with $\nu_{\rm LF}({\rm max}) \approx \nu_{\rm nod}({\rm ISCO})$,
providing a very good estimate for the system parameters. We see that
the upper limit on the mass of XTE J1550-564 obtained from this method
is very close to the dynamical measurement of $M=9.1$ (\citealt{Orosz2011}).\\

\subsection{The case of H1743-322}

\begin{figure}
\begin{centering}
 \includegraphics[height=8cm,width=8cm,trim=0.0cm 0.0cm 0.0cm 0.0cm,clip=true]{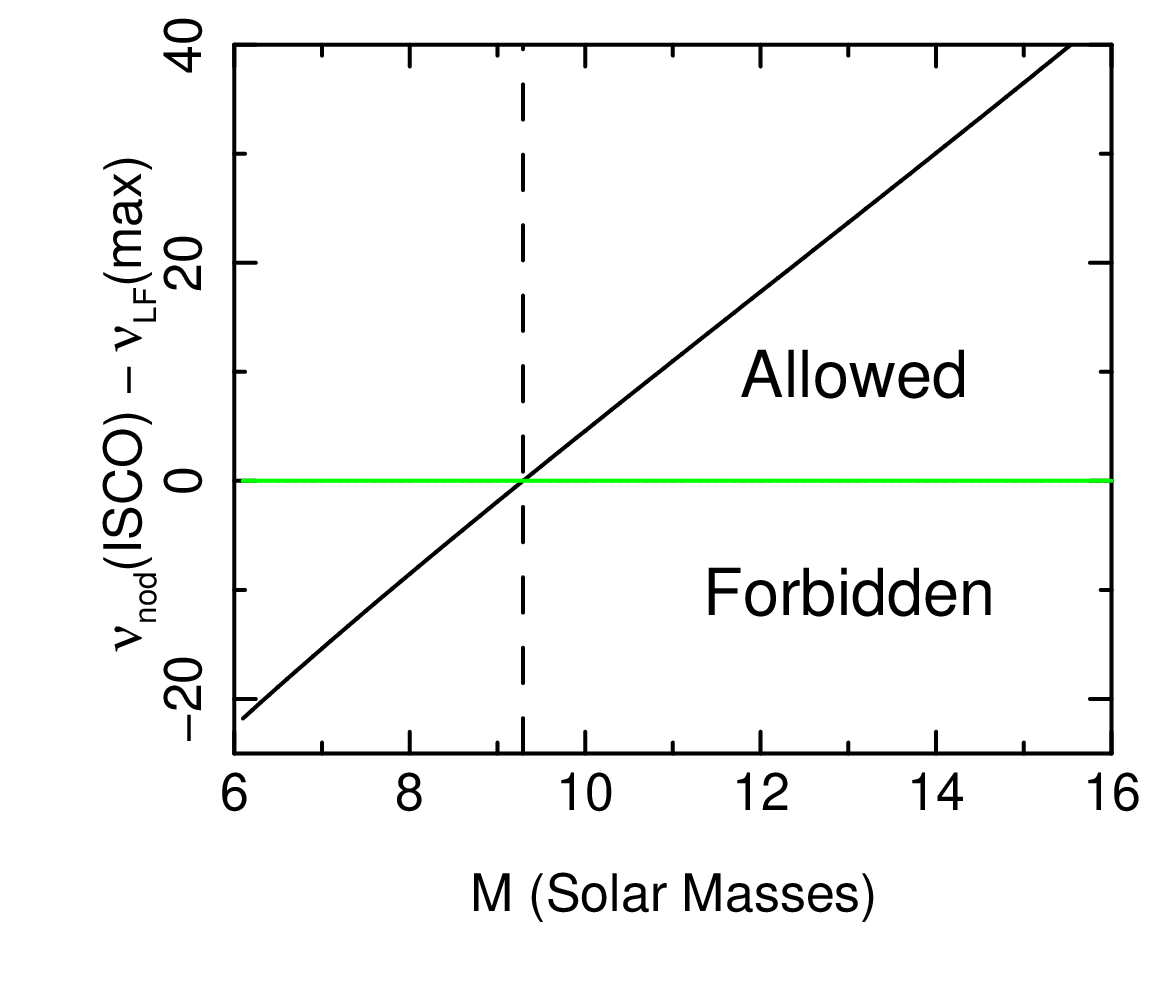}
\end{centering}
\vspace{-0.5cm}
 \caption{Function $f(\nu_{\rm \phi} , \nu_{\rm per} ,\nu_{\rm nod} , \nu_{\rm LF}({\rm max}) ) =
\nu_{\rm nod}({\rm ISCO}) - \nu_{\rm LF}({\rm max})$ for the case of H
1743-322, when only $\nu_{\phi}$ and $\nu_{\rm per}$ are known. The
intersection between the black line and the green line (that marks
$\nu_{\rm nod}({\rm ISCO}) - \nu_{\rm LF}({\rm max}) = $ 0)
corresponds to the lower limit on $M$ ($M\geq 9.29$).}
 \label{fig:ISCOtest_1743}
\end{figure}

It is easy to see that this procedure will work when $\nu_{\rm nod}$
is the unknown instead of $\nu_\phi$, as is the case for an
observation of H1743-322 (\citealt{Homan2005}). We note that this is
unusual, since the Type-C QPO is far easier to detect than the HF QPOs
but, evidently, not impossible. In this case, the HF QPOs have
frequencies $\nu_\phi = 204$ Hz and $\nu_{\rm per} = 165$ Hz, plus the
highest detected Type-C QPO frequency is $\nu_{\rm LF}({\rm
  max})=9.44$ Hz (see Table \ref{tab:solutions}). We can calculate limits on
$M$, $a$ and $r$ by making guesses for $M$. For each $M$ trial value,
we calculate the corresponding $r$ by solving equation \ref{eqn:x2}
and use this to calculate $a$ from equation \ref{eqn:Sigma}. From
this, the ISCO can be calculated and, finally, the nodal frequency at
the ISCO. In Figure \ref{fig:ISCOtest_1743}, we plot the resulting
function $f = \nu_{\rm nod}({\rm ISCO}) - \nu_{\rm LF}({\rm max})$
against the trial value of $M$. Since this function must be positive
if there are to be no orbits inside the ISCO, the mass must be to the
right of the dotted line. Again using a bisection search, we obtain
the limits $M \geq 9.29$, $a \geq 0.21$ and $r \leq 5.89$.

In the final case where $\nu_{\rm per}$ is the unknown frequency,
limits can be placed by calculating the same function with trial
values of $\nu_{\rm per}$ and finding its root.

%%Left, Bottom, Right, Top

%\twocolumn  																				
%---------------------------------------------------------
\section{Discussion \& Conclusions}
\label{sec:conclusions}
%---------------------------------------------------------

We have presented analytic / inexpensive numerical solutions to the
RPM equations. This paper is primarily intended as a `cookbook' for
measuring mass and spin using the RPM and, to that end, we have
written a user friendly \textsc{fortran} code which finds solutions,
with error estimates, for any of the cases mentioned here. Any
interested readers wishing to use this code are encouraged to contact
us. 

For the case when three simultaneous QPOs are observed, we have
found analytical solutions to derive the mass and spin of the black
hole. For the case when only two QPOs are detected simultaneously, a
dynamical mass measurement can be combined with the QPO frequencies to
provide a spin measurement. Even when no dynamical mass measurement
exists, we can still place limits on the spin of the black hole by
requiring that the highest Type-C QPO frequency ever observed from the
source must come from an orbit larger than or coincident to the ISCO. 

We note that, in
principle, we could also solve for mass with two simultaneously
detected QPOs and a measurement of spin via spectroscopic methods
(i.e. fitting the disk spectrum or iron line profile;
\citealt{Kolehmainen2010}; \citealt{Fabian2012}). However,
the large uncertainties associated with the spin, particularly after
comparison between the disk and iron line estimates, limit the
usefulness of this exercise.\\

\noindent We note that the RPM simply considers test mass orbits in the Kerr
metric. Further theoretical framework is required to understand
exactly how these frequencies will modulate the X-ray flux from an
accretion flow, which comprises an optically thick, geometrically thin
accretion disk (e.g. \citealt{Shakura1973}) and some optically thin
electron cloud emitting a Comptonised power law spectrum
(\citealt{Thorne1975}). The LF QPO model of \cite{Ingram2009}
considers a truncated disk / inner hot flow geometry in which nodal
precession of the entire inner flow results from the frame dragging
effect. This naturally explains how the precession frequency modulates
the X-ray flux (\citealt{Ingram2012a}; \citealt{Veledina2013}) and how
a coherent LF QPO can be observed even when the inner flow is thought
to be rather extended. The HF QPOs, on the other hand, are only
observed with high frequencies, when the disk truncation radius is
thought to be close to the ISCO. We do, however, see broad power
spectral features with characteristic frequencies that co-evolve with
the LF QPO frequency and apear to eventually evolve into HF QPOs
(\citealt{Psaltis1999}). M14 and M14a used their spin and mass
constraints for GRO 1655-40 and XTE J1550-564 respectively to
demonstrate that this co-evolution is roughly consistent with the high
frequency features peaking at $\nu_{\rm per}$ and $\nu_\phi$ and the
LF QPO peaking at $\nu_{\rm nod}$, all for a moving radius. 

Thus, perhaps the periastron precession and orbital frequencies
modulate the X-ray flux through randomly occurring anisotropies in the
inner accretion flow (see e.g. \citealt{Wellons2014};
\citealt{Schnittman2006}). \cite{Bakala2014} demonstrate that HF QPOs
at the epicyclic frequencies are seen from X-ray emitting 
blobs orbiting a BH, resulting mainly from variable Doppler
effects. When the inner flow is extended  (and assuming that
modulations from different radii are not correlated; see
e.g. \citealt{Ingram2013}), there will be modulations at a large range
of frequencies, resulting in broad power spectral features. When the
truncation radius is close to the ISCO, this picture converges to the
RPM in which the anisotropies only occur for a thin ring, and so we
see QPOs at three frequencies. We will develop a formalism for this
model in a future paper in order to test it against the observed
co-evolution of the broad high frequency features and the LF QPO.

For the simultaneously detected QPOs considered here, there are two
objects for which we are able to make multiple spin measurements. For
GRO 1655-40, we obtain 3 spin estimates which are all consistent with
one another. For XTE J1550-564, we obtain an upper limit of $a \leq
0.341$, consistent with the measurement of $a = 0.339 \pm 0.007$
presented here and in M14a. This is encouraging, and we note that the
\textit{Large Observatory For x-ray Timing} (\textit{LOFT};
\citealt{Feroci2011}), should it fly, will detect many more triplets of HF QPOs to
test the RPM more thoroughly.

\section*{Acknowledgments}
AI and SEM acknowledge the Observatory of Rome for hospitality. SEM
acknowledges support from the ESA research fellowship program.

\bibliographystyle{mn2e}
\bibliography{biblio} 

\onecolumn

\begin{center}
\begin{landscape}																				
\begin{longtable}{c c c c c c | c c c}                                                                                                                                                                     																				
\caption{Solutions of the RPM obtained through the analytical and semi-analytical methods presented in this work. The frequencies have been taken from the litterature. }\label{tab:solutions} \\        																				
																				
\hline																				
\multicolumn{1}{c}{} & \multicolumn{1}{c}{Type-C} & \multicolumn{1}{c}{Lower HF QPO} & \multicolumn{1}{c}{Upper HF QPO} & \multicolumn{1}{c}{Mass} & \multicolumn{1}{c}{Ref.} & \multicolumn{3}{|c}{Solutions}\\                                                                                                                                                                                                                                                           																				
		&	Frequency [Hz]	&	Frequency [Hz]	&	Frequency [Hz]	&		&				&	Mass	&	Spin	&	Radius	\\
	\hline	
	\hline																		
	\multicolumn{9}{c}{{\bf GRO J1655-40}}\\																			
	\hline																			
	\hline																			
	3 QPOs	&	$ 17.3 _{ -0.1 }^{+0.1 }$	&	$298_{-4}^{+4}$	&	$441_{-2}^{+2}$	&		&	(1)			&	5.31 $\pm$ 0.07  M$_{\odot}$	&	0.285 $\pm$ 0.003	&	5.68 $\pm$ 0.04	\\

	2 QPOs and mass	&	$18.3_{-0.1}^{+0.1}$	&		&	$451_{-5}^{+6}$	&	5.4 $\pm$ 0.3 M$_{\odot}$	&	(1, 2)			&		&	0.28 $\pm$ 0.02	&	5.5 $\pm$ 0.2	\\
	2 QPOs and mass	&	$18.1_{-0.1}^{+0.1}$	&		&	$446_{-4}^{+4}$	&		&				&		&	0.29  $\pm$ 0.02	&	5.6 $\pm$ 0.2	\\
																				
	\hline																			
	\hline																			
	\multicolumn{9}{c}{{\bf XTE J1550-564}}\\																			
	\hline																			
	\hline																			
	2 QPOs and mass	&	13.08   $\pm$0.08       	&	183     $\pm$5	&		&	9.10 $\pm$ 0.61 M$_{\odot}$	&	(3, 4)			&		&	0.339 $\pm$ 0.007	&	5.47$\pm$0.12	\\
																				
	2 QPOs	&	13.08   $\pm$0.08       	&	183     $\pm$5	&		&		&	(3)			&	$\leq$9.56 M$_{\odot}$	&	$\leq$ 0.341	&	$\geq$ 5.39	\\
		&	highest detected: 18.04 $\pm$ 0.07	&		&		&		&				&		&		&		\\
	\hline																			
	\hline																			
	\multicolumn{9}{c}{{\bf H1743-322}}\\																			
	\hline																			
	\hline																			
	2 QPOs	&	highest detected: 9.44 $\pm$ 0.02	&	165 $^{+9}_{-5}$	&	240 $\pm$ 3	&		&	(5)			&	$\geq 9.29$M$_{\odot}$	&	$\geq 0.21$	&	$\leq 5.89$	\\
																				
	\hline																			
																				
\end{longtable}	
%\begin{flushleft}																	
\textsc{REFERENCES:} 																	
	(1) \cite{Motta2014}; (2) \cite{Beer2002} (3) \cite{Motta2014a}; (4) \cite{Orosz2011}; (5) \cite{Homan2005}. 																
%\end{flushleft}																							
\end{landscape}                                                                                                                                                                                                                                																				
\end{center}

\label{lastpage}
\end{document}